\documentstyle[aps,twocolumn,epsfig]{revtex}

\begin{document}
%\draft
\title{SINGULARITY FORMATION AND COLLAPSE IN THE
ATTRACTIVE GROSS-PITAEVSKII EQUATION}
\author{A.V. Rybin${}^{\dag}$, G.G. Varzugin${}^{\ast}$ and J.
Timonen${}^{\dag}$}
\address{{\dag Department of Physics, University of Jyv{\"a}skyl{\"a}}\\
 {PO Box 35, FIN-40351}
{Jyv{\"a}skyl{\"a}, Finland}}
\address{{$\ast$ Institute of Physics}
 {St. Petersburg State University}\\
 {198904, St. Petersburg, Russia}}

\maketitle

\begin{abstract}

A generic mechanism of collapse in the Gross-Pitaevskii equation
with attractive interparticle interactions is gained by
reformulating this equation as Newton's equation of motion for a
system of particles with a constraint. 'Quantum pressure' effects
give rise to formation of a potential barrier around the
emerging singularity, which prevents a fraction of the particles from
falling into the singularity. For reasonable initial widths of the
condensate, the fraction of collapsing particles, which are thereby
removed from the condensate, is found to be a 'universal' number
$\simeq 0.7$.
\end{abstract}

\pacs{05.30.-d, 05.45.-a,03.75.Fi}

The Bose-Einstein condensate formed  in a  magneto-optically
trapped vapor of cold alkali atoms,  can be quite  accurately
described by the Gross-Pitaevskii (GP) equation~\cite{pit}
\begin{equation}
i \hbar\Psi_t + \frac{\hbar^2}{2m}\Delta\Psi-g\Psi|\Psi|^2-
V(\bbox{ x})\Psi=0,\label{GP}
\end{equation}
where $\Psi(\bbox{ x},t)$ is the wave function of the condensate,
the external potential $V(\bbox{ x})$ models the wall-less
confinement (the trap), $m$ is the mass of an individual atom,
$g=4\pi\hbar^2m^{-1}a_s$ is the effective interaction strength
(the 'coupling constant') with $a_s$
 the scattering length, and
$\Delta=\sum_i\frac{\partial^2}{\partial x_i^2}$ is the Laplace
operator. A convenient as well as practical choice for the confining
trap is the paraboloidal potential $V=\frac{m}{2}\sum_i^3\omega_i^2
{x_i}^2$.

It has been known~\cite{pit,x,xx,xxx,xxxx,xxxxx,xxxxxx} for some
while already that the Bose-Einstein condensate of atoms with a
negative scattering length becomes unstable when the number of
atoms exceeds some critical value $N_c$. These instabilities have
recently been analyzed~\cite{sackett} experimentally. To this end
the most interesting prospects are provided by the
application~\cite{cornish} of Feshbach resonance by which one can
actually tune the effective scattering length with an external
magnetic field, and thereby e.g. sweep the coupling constant from
positive to negative. This technique was very recently
used~\cite{sackett} on a condensed gas of ${}^{85}\mbox{Rb}$
atoms such that a collapse and the consecutive 'explosion' of a
condensate of ${}^{85}\mbox{Rb}$ atoms was created.

Triggered by these recent experiments, a possible mechanism
through elastic collisions for the ejection from the condensate of
atoms with negative scattering length was discussed
in~\cite{stoof}. In this Letter we show however  that the
mean-field theory (i.e. the GP equation) of the condensate already
allows for a description  of 'partial collapse', and leads to an
accurate estimate for the number of particles that remain in the
condensate after the explosion. The fraction of the condensate
thrown away in the collapse (i.e. the collapsing part of the
condensate), is always about 70$\%$ provided that the width of the
condensate before the change of sign of the scattering length is
close to that of the ground state. This was true e.g. for the
experiment reported in~\cite{cornish}. We find that the collapsing
fraction is relatively insensitive to the total number of
particles in the condensate, and that this fraction tends to
100$\%$ only in the limit of very large width of the condensate.

Collapse phenomena are very well known in nonlinear optics and
plasma physics, and they have there a long history extending over
the last thirty years (see e.g. review~\cite{rassmussen} and
references therein). Following this tradition, the collapse in the
GP system with a negative scattering length was discussed in a
similar fashion in~\cite{pitc,wad,berge}.

Within this tradition, the analysis of  singularity formation is
mainly focussed  on the   self-similar behaviour of the solutions
in the vicinity of the singularity.  We have recently shown
~\cite{rybin}, however, that a true self-similar solution of the
GP system is only possible in  two space dimensions. This means
that a self-similar description of collapse can only be used with
certain reservations. In this Letter we propose a completely
different description for the singularity formation, which does
not rely on self-similarity of the solution. In a sense the
starting point of our analysis is the concept of 'partial
collapse', now of the Bose-Einstein condensate. A suggestion about
a singularity restricted to a localized domain inside the
condensate, was already made in~\cite{ueda}. Here we elaborate
further on this suggestion, and formulate it within a completely
different framework  from that used in ~\cite{ueda}.

Our approach is based on the  realization that the GP equation
Eq.~(\ref{GP}) can be interpreted as Newton's equations of motion
for (fictitious) particles with a certain constraint. The dynamics
of GP collapse can then easily be followed by following the
trajectories of these particles, under the influence of an
effective potential due to 'pressure' effect that arise from the
spatially varying condensate density. By introducing an auxiliary
field $\bbox{\eta}(\bbox{ x},t)$, and its inverse $\bbox{
r}(\bbox{\eta},t)$, we can show (see below) that these
trajectories satisfy a system of ordinary differential equations,
\begin{equation}
\frac{\partial r_i}{\partial
t}=\frac{1}{m}\frac{\partial\phi}{\partial x_i}(\bbox{
r}(\bbox{\eta},t),t);\;\;\;r_i(\bbox{\eta},0)=\eta_i,\label{Eqforr}
\end{equation}
with $i=1,2,3$ for d=3 dimensions. Here  $\phi$ is a phase defined
by the representation $\Psi=\sqrt{\rho} e^{i\phi/\hbar}$ of the
condensate field. The fields $r_i(\bbox{\eta},t)$ can now be
interpreted as trajectories of some particles with initial points
$\eta_i$. We describe in the following how the GP equation
Eq.~(\ref{GP}) can be transformed into the equations of motion for
this system of particles.

It is plain that, by using $\Psi=\sqrt{\rho} e^{i\phi/\hbar}$,
Eq.~(\ref{GP}) is equivalent to a coupled set of equations for
$\rho$ and $\phi$,

\begin{equation}
\rho_t+\frac{1}{m}\mbox{div}(\rho\nabla\phi)=0\label{Eqcontinuity}
\end{equation}
\begin{equation}
\phi_t+\frac{1}{2m}(\nabla\phi)^2+V+
g\rho-\frac{\hbar^2}{2m}\rho^{-1/2}\Delta\rho^{1/2}=0.
\label{Eqphase}
\end{equation}

Let us now introduce an auxiliary field $\bbox{\eta}(\bbox{
x},t)$, which is assumed to satisfy

\begin{equation}
\frac{\partial\eta_i}{\partial
t}+\frac{1}{m}\nabla\eta_i\nabla\phi=0.\label{Eqforeta}
\end{equation}
If we define a Jacobian $$J(\bbox{ x},t)=\det
J^m_n,\;\;\;J^m_n=\frac{\partial\eta_m}{\partial x_n},$$ it then
follows from Eq.~(\ref{Eqforeta}) that

\begin{equation}
\frac{\partial\ln J}{\partial t}+\frac{1}{m}\nabla\ln
J\nabla\phi+\frac{1}{m}\Delta\phi=0. \label{eqforg}
\end{equation}
From this  Eq.~(\ref{eqforg}) we can easily deduce that

\begin{equation}
\rho(\bbox{ x},t)=J(\bbox{ x},t)\rho_0(\bbox{\eta})\label{ansatz}
\end{equation}
is a solution of the continuity equation Eq.~(\ref{Eqcontinuity})
with $\rho_0$  an arbitrary function. We now define the field
$\bbox{ r}(\bbox{\eta},t)$ by  requiring  that the function
$\bbox{ r}(\bbox{\eta},t)$ is the inverse of the function
$\bbox{\eta}(\bbox{ x},t)$. With this definition
Eq.~(\ref{Eqforeta}) and Eq.~(\ref{Eqforr}) become equivalent.
Notice that the relation Eq.~(\ref{ansatz}) provides the
anticipated representation of the condensate density in terms of
particle trajectories.

Thus far we have not used  Eq.~(\ref{Eqphase}). By differentiating
Eq.~(\ref{Eqforr}) with respect to $t$, we find that

\begin{equation}\label{8}
m\frac{\partial^2r_i}{\partial t^2}=\frac{\partial}{\partial
x_i}(\phi_t+\frac{1}{2m}(\nabla\phi)^2).
\end{equation}
Using now Eq. (\ref{Eqphase}), this Eq.~(\ref{8}) can be expressed
in the familiar form of Newton's equations of motion for a set of
particles,

\begin{equation}
m\frac{\partial^2r_i}{\partial t^2}=-\frac{\partial V}{\partial
x_i}-\frac{\partial U}{\partial x_i}, \label{EqNewton}
\end{equation}
in which the 'pressure potential' $U(\bbox{ x},t)$ is

\begin{equation}
\label{pressures} U(\bbox{
x},t)=-\frac{\hbar^2}{2m}\rho^{-1/2}\Delta\rho^{1/2}+ g\rho.
\end{equation}
The 'particles' of this system thus feel, in addition to the
confining trap potential, an effective potential that arises from
the spatially and temporally   varying condensate density
$\rho(\bbox{ x},t)$. The second term in the right side of
Eq.~(\ref{pressures}) can be recognized as the conventional
pressure term, while the first term there arises from the kinetic
energy part of the GP equation, and can be considered as a quantum
pressure term. As will be explained below, it is the interplay
between these two effects in the pressure potential which
determines the dynamics of the collapse.

As can be readily seen, the result  Eq.~(\ref{ansatz}) can also be
written in the form

\begin{equation}
\det\left(\frac{\partial r_i}{\partial\eta_k}\right)\rho(\bbox{
r},t)=\rho_0(\bbox{\eta}). \label{relation}
\end{equation}
This constraint  can now be used to exclude the particle density
$\rho$ from Eq.~(\ref{EqNewton}). This means that the system of
equations Eqs.~(\ref{EqNewton}-\ref{relation}) is self consistent,
and fully describes the dynamics of the condensate. We need
however to specify first the initial conditions for this system,
and a natural choice for Eq.~(\ref{EqNewton}) is

\begin{equation}
r_i(\bbox{\eta},0)=\eta_i,\;\;\;\frac{\partial r_i}{\partial
t}(\bbox{\eta},0)=v_i(\bbox{\eta}),
\end{equation}
with $\bbox{ v}=\frac{1}{m}\nabla\phi|_{t=0}$. The function
$\rho_0$ can now be interpreted as the initial density of the
particles, and the phase $\phi$ can be solved from
Eq.~(\ref{Eqforr}) up to a time-dependent factor.

Notice that there is a relation between $\phi$ and $\rho$. Indeed,
from Eq.~(\ref{Eqforr}) we find that $$ \det\left(\frac{\partial
r_i}{\partial
\eta_k}\right)=e^{\frac{1}{m}\int_0^t\Delta\phi(\bbox{ r},t')dt'}.
$$ Hence, we can express $\rho$ in the form
\begin{equation}
\rho(\bbox{ x},t)=e^{-\frac{1}{m}\int_0^t\Delta\phi(\bbox{
x}(t,t'),t')dt'}\rho_0(\bbox{\eta}(\bbox{ x},t)),
\end{equation}
where $\bbox{ x}(t,t^\prime)=\bbox{ r}(\bbox{\eta}(\bbox{
x},t),t^\prime).$

It is evident that a possible collapse of the condensate is
related to the singular points of the solution,  i.e. to the
points where the function $\bbox{ r}(\bbox{\eta},t)$ is not
invertible and the determinant $\det(\frac{\partial
r_i}{\partial\eta_k})$ vanishes. At these points the density
becomes infinite. It seems natural to assume that if collapse
takes place, it happens at a finite time $t=t_\ast$, {\it i.e.}
$\rho(\bbox{ r}_s,t_\ast)=\infty$. Then, from
Eq.~(\ref{relation}), we can conclude that for any $\bbox{ r}_s$
there is a set $\Omega_s$ such that $$ \lim_{t\to
t_\ast}r_i(\bbox{\eta},t)=r_{si} $$ for all
$\bbox{\eta}\in\Omega_s$. In other words, all particle
trajectories that begin at any $\bbox{\eta}\in\Omega_s$ are
converged  at one point $\bbox{ r}_s$.

Suppose that the set of singular points, $\{\bbox{ r}_s\}$,
consists of one isolated point while $\Omega_s$ is a domain. We
can then define the number of particles involved in the collapse
as

\begin{equation}
N_e=\int_{\Omega_s}\rho_0(\bbox{\eta})d^3\eta.\label{ne}
\end{equation}
This number can thus be interpreted as  the number of particles
that are  ejected from the condensate when it is collapsed. Notice
that the domain $\Omega_s$ can also coincide with  the whole
space. In this case $N_e=N$, the total number of atoms in the
condensate. This is true in particular for the self-similar
solutions, when  the whole condensate collapses to a
$\delta$-function centered in the trap, and  completely disappears
apart from the very particular borderline case of an oscillating
solution~\cite{rybin}. The self-similar solutions can only
exist~\cite{rybin} in two space dimensions. It is important to
emphasize  that, {\it in general}, $N_e<N$. It can be shown, {\it
e.g. }, that if $\rho_0$ tends to a Gaussian distribution as
$|\bbox{\eta}|\to\infty$, then $N_e<N$. This means that $N_e$ is
an essential  characteristics  of the collapse of  attractive
condensates.

We can  explain the partial collapse qualitatively as follows. If,
initially, the width of the condensate is greater than the
characteristic  oscillator length  $a_{HO}$, both the quantum
pressure and the conventional pressure induce the condensate to
decrease its width. However, when the width becomes equal or less
than $a_{HO}$, the two pressure terms  in the energy have opposite
effects: quantum pressure favours an expanding condensate (as a
result of an 'uncertainty principle'), while the negative
conventional pressure term favours a collapsing condensate.
Provided that the number of particles is greater than a critical
value, the negative pressure dominates over the quantum pressure
in a small domain centered at a point $\bbox{ r}_s$.  Outside this
domain it is the quantum pressure that dominates, and prevents the
outside particles from entering this domain.

In order to estimate the value of $N_e$,  we use here a Gaussian
trial wave function\cite{y,yy,yyy,yyyy}. It gives an approximate
solution for the density, which can be expressed in the form

\begin{equation}
\rho(\bbox{ x},t)=\frac{N}{\pi^{3/2}\prod_i\tau_i(t)}
e^{-\sum_i\frac{x_i^2}{\tau_i^2(t)}}, \label{Eqlocalrho2}
\end{equation}
in which the functions $\tau_i$ satisfy~\cite{gar,kagan,dalfovo}

\begin{equation}
\ddot{\tau}_i+\omega_i^2\tau_i-\frac{\hbar^2}{m^2}\frac{1}{\tau_i^3}-
\frac{gN}{(2\pi)^{3/2}m}\frac{1}{\prod_k\tau_k}\frac{1}{\tau_i}=0
\label{Eqonsigma}
\end{equation}
with $\tau_i(0)=a_i$ and $\dot{\tau}(0)=0$. Here $a_i$ is the
initial width of the condensate.

Substituting the solution Eq.~(\ref{Eqlocalrho2}) in
Eq.~(\ref{EqNewton}), we find that

\begin{eqnarray}
\frac{\partial^2r_i}{\partial t^2}&=&-\frac{\partial
V_{eff}}{\partial r_i}\label{Eqeffective} \\
V_{eff}(\bbox{r},t)&=&\frac{1}{2}\sum_i\left(\omega_i^2
-\frac{\hbar^2}{m^2\tau_i^4}\right)r_i^2
+\frac{g}{m}\rho(\bbox{r},t).\nonumber
\end{eqnarray}
Here $V_{eff}$ is an effective time-dependent potential that
determines the behaviour of particle trajectories in the vicinity
of the center of the trap.

For simplicity, we consider here in more detail the symmetric
trap: $\omega_i=\omega$ and $a_i=a$. Hereafter we measure the time in
units of $1/\omega$,  and the distance in units of
$a_{HO}=\sqrt{\hbar/m\omega}$,   the oscillator length. Now
Eqs.~(\ref{Eqeffective}),(\ref{Eqonsigma}) take the form

\begin{eqnarray}
\frac{d^2r}{dt^2}&=&-\frac{\partial V_{eff}}{\partial r}\label{Eqsymmodel}\\
V_{eff}&=&\frac{1}{2}\left(1-\frac{1}{\tau^4}\right)r^2
-\frac{4}{\sqrt{\pi}}\frac{g_0}{\tau^3}
\left(e^{-\frac{r^2}{\tau^2}}-1\right),\nonumber
\end{eqnarray}
and

\begin{equation}
\frac{d^2\tau}{dt^2}+\tau-\frac{1}{\tau^3}+
\sqrt{\frac{2}{\pi}}\frac{g_0}{\tau^4}=0,
\label{tau}
\end{equation}
in which $g_0=\frac{|a_s|N}{a_{HO}}$, and the initial conditions
are given by $r(0)=\eta, \dot r(0)=0, \tau(0)=a, \dot\tau(0)=0.$
Notice that if $g_0>g_{0cr}$, where $g_{0cr}\approx
0.671$~\cite{pit}, the solution of Eq.~(\ref{tau}) has a zero at a
finite time $t=t_\ast$.

\begin{figure}
\epsfysize = 200pt
\centerline{\epsfbox{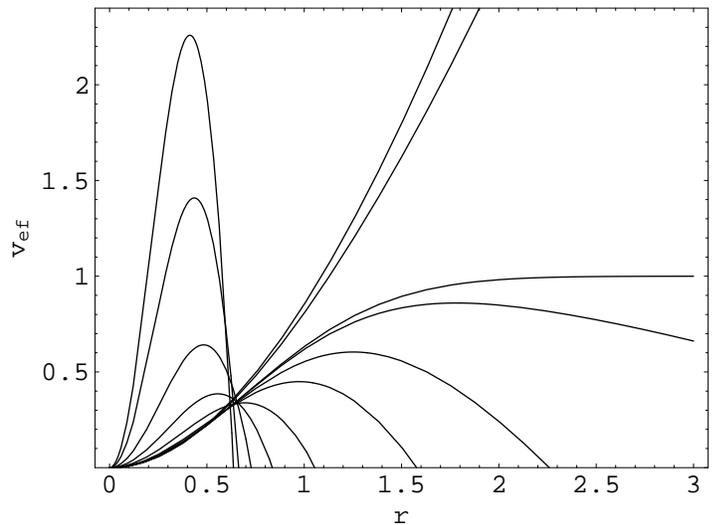}}
\caption{The effective potential $V_{eff}$ at different times
$t<t_\ast$. The initially almost paraboloidal potential develops a
growing barrier at the final stages of the collapse.
\label{Fig.1}}
\end{figure}

Let us assume that the system has $a>1$ and begins to collapse.
Initially, $V_{eff}$ can be well approximated by a paraboloidal
trap. The particles are  drifted towards the center and the
condensate width decreases. When this width is reduced to the
oscillator length, a potential barrier is however formed at a
distance $r=r_{max}$ from the center,
$$r_{max}=\tau\sqrt{\ln\frac{8g_0}{\sqrt{\pi}(\tau-\tau^5)}},$$
preventing the outside particles from falling into the emerging
singularity. Figure 1 illustrates the  formation of this potential
barrier in the limit when the width of the condensate $\tau$ goes
to zero.

We solved the system Eqs.~(\ref{Eqsymmodel}),(\ref{tau})
numerically. The collapse time was obtained from the
condition\footnote{The final stage of the collapse takes place
very fast, so a good numerical estimate for $t_\ast$ was already
obtained from this seemingly rather crude approximation, which
however allowed us to avoid the difficulties related to the
mathematical singularity at $t=t_\ast$.} $\tau(t_\ast)<0.1$.
Equation~(\ref{Eqsymmodel}) was solved in the time interval
$[0,t_\ast]$, while the point $\eta$ was supposed to belong to the
domain $\Omega_s$ provided that $r(t_\ast)<r_{max}(t_\ast)$.

The conditions of the experiment reported in~\cite{cornish} were
modelled such that the initial width of the condensate was derived
as a stationary solution of Eq.(\ref{tau}) with $g_0$ replaced by
$-41.67g_0$. Notice that in the experiment~\cite{cornish} the
initial positive scattering length was $2500a_0$ ($a_0$ is the
Bohr radius), and this scattering length was in $0.5\mbox{ms}$
changed to  $-60 a_0$. The number $-41.67$ is thus the ratio of
these two scattering lengths. We varied the parameter $g_0$  in a
large range, from 1 to 300. In all cases the fraction of the
particles ejected from the condensate as a result of collapse, was
about $70\%$. We also found that the fraction of ejected particles
increased above $70\%$ only when the initial width of the
condensate was increased well above that of the ground state. In
the limit of very large widths the fraction approached $100\%$ of
the total number of particles.

In summary, in this Letter we have explained the generic mechanism
of collapse in a trapped Bose-Einstein condensate with attractive
interparticle interactions   in the framework of mean-field theory
(GP equation). We can conclude that, in general, only a fraction
of the condensate takes part in the collapse, while the rest of
the particles are screened from the emerging singularity by an
effective potential barrier due to 'quantum pressure' effects.
This potential barrier is somewhat reminiscent of the 'event
horizon' that appears around gravitational singularities. We could
also show that, for reasonable initial widths of the condensate,
the fraction of particles removed from the condensate as a result
of collapse is an almost 'universal' number $\simeq 0.7$. Even
though we here derived this fraction of ejected particles only
approximately, our definition of the related particle  number
$N_e$ Eq.~(\ref{ne}), can be made mathematically rigorous. This
number can be obtained from the solution of Eq.~(\ref{EqNewton}).
The simplified   model based on Gaussian trial wave function,
which we  proposed here, is not accurate if the 'coupling
constant' $g_0$ is in the vicinity of $g_{0cr}$. A more thorough
analysis of Eq.~(\ref{EqNewton}) will be reported in a forthcoming
publication.

\acknowledgments This work was supported by the Academy of Finland
under the Finnish Center of Excellence Programme 2000-2005
(Project No. 44875, Nuclear and Condensed Matter Physics at JYFL).
GGV was partly supported by an RFBR grant No. 00-01-00480.

%\begin{figure}
%\caption{The effective potential $V_{eff}$ at different times
%$t<t_\ast$. The initially almost paraboloidal potential develops a
%growing barrier at the final stages of the collapse.
%\label{Fig.1}}
%\end{figure}

\end{document}